\documentclass[12pt,reqno]{amsart}
\usepackage{graphicx}
\usepackage{amscd}
\usepackage{amsmath}
\usepackage{epsfig}
\usepackage{amsfonts}
\usepackage{amssymb}

\usepackage{hyperref}

\providecommand{\U}[1]{\protect\rule{.1in}{.1in}}
\providecommand{\U}[1]{\protect\rule{.1in}{.1in}}
\allowdisplaybreaks

\theoremstyle{plain}

\numberwithin{equation}{section}

\date{June 21, 2021}

\begin{document}
\title[Relativistic Schr\"{o}dinger equation]{Discovery of the Relativistic
Schr\"{o}dinger equation\/}
\author{Kamal Barley}
\address{Department of Applied Mathematics and Statistics, Stony Brook University, NY
11794-3600, U.~S.~A.}
\author{Jos{\'{e}} Vega-Guzm{\'{a}}n}
\address{Department of Mathematics, Lamar University, 200 Lucas Building, P.~O. Box 10047, Beaumont, TX
77706, U.~S.~A.}
\author{Andreas Ruffing}
\address{Landeshauptstadt M\"{u}nchen, Referat f\"{u}r Bildung und Sport,
Seminar TMG Universit\"{a}t, Drygalski Allee~2,
81477 M\"{u}nchen, Germany 11794-3600}
\author{Sergei K. Suslov}
\address{School of Mathematical and Statistical Sciences, Arizona State University, P.~O. Box 871804, Tempe, AZ 85287-1804,
U.~S.~A.}

\begin{abstract}
We discuss the discovery of the relativistic wave equation for a spin-zero
charged particle in the Coulomb field by Erwin Schr\"{o}dinger (and elaborate on why he didn't publish it).
\medskip

Wir beleuchten die Entdeckung der relativistischen
Wellengleichung f\"{u}r ein geladenes Spin-Null-Teilchen
im Coulomb-Feld durch Erwin Schr\"{o}dinger (und
erarbeiten, warum er diese Gleichung nicht
ver\"{o}ffentlicht hat).
\end{abstract}

\dedicatory{{One cannot escape the feeling that these mathematical formulae have an independent existence and an intelligence of their own, that they are wiser than we are, wiser even than their discoverers, that we get more out of them than
was originally put into them.
\newline Heinrich Hertz, on Maxwell's equations of electromagnetism\/}
\medskip
\newline{Man kann nicht der Empfindung entrinnen, dass
diese mathematischen Formeln eine unabh\"{a}ngige
Existenz und eigene Intelligenz besitzen, dass sie
weiser als wir sind, weiser sogar als ihre Entdecker,
dass wir mehr aus ihnen gewinnen, als das, was
wir an Einsatz f\"{u}r sie urspr\"{u}nglich erbracht haben.
\newline
Heinrich Hertz, \"{U}ber Maxwells Gleichungen des Elektromagnetismus\/}}

\maketitle

\section{Introduction\/}

There are many great reminiscences and reviews of remarkable events of the early days of quantum
mechanics. Among those are \cite{Bloch1976}, \cite{Blokhintsev1977}, \cite{Dirac1979}, \cite{Eistein2011}, \cite{Elyashevich1977}, \cite{Joas2009}, \cite{Kragh1982}, \cite%
{Mehra1976}, \cite{Mehra1987a}, \cite{Mehra2001}, \cite{Mehra1987}, \cite{Milantev2004}, \cite{Miller2002}, \cite{Renn2013}, \cite{Smorodinskii1988}, \cite{Waerden1968}, \cite{Wessels1979} (and references therein), as well as great biographies of the `pioneers' of the quantum
revolution, namely, Werner Heisenberg \cite{Cassidy1991}, \cite{Cassidy2009}, Erwin Schr\"{o}dinger
\cite{Gribbin2012}, \cite{Moore1989}, and Paul Dirac \cite{Farmelo2009}.

A traditional story of the discovery of Schr\"{o}dinger's wave equation -- \textquotedblleft rumor has it\textquotedblright --
goes this way \cite{Bloch1976}, \cite{Gribbin2012}, \cite{Moore1989}: In early November 1925, Peter
Debye, at the ETH,%
\footnote{Swiss Federal Institute of Technology in Z\"{u}rich.}
asked Schr\"{o}dinger to prepare a talk for the Z\"{u}%
rich physicists about de Broglie's work just published in\textit{\ Annales
de Physique.} The date on which this particular colloquium took place has
not been recorded, but it must have been in late November or early December,
before the end of the academic term \cite{Gribbin2012}, \cite{Mehra1987}. According to Felix Bloch
\cite{Bloch1976},%
\footnote{An alternative recollection is due to Kapitza \cite{Kapitza1980}.}
at the end of Schr\"{o}dinger's talk, Debye casually remarked
that (de Broglie's) way of talking was rather childish and someone had to
think about an equation for these (de Broglie's) `matter
waves'. As a student of Sommerfeld, Debye had learned
that to deal properly with waves, one had to have a wave equation. Indeed,
at his second talk, just a few weeks later, Schr\"{o}dinger mentioned:
\textquotedblleft My colleague Debye suggested that one should have a wave
equation; well, I have found one!\textquotedblright

For a sequence of those remarkable events, a direct quote is quite appropriate \cite%
{Gribbin2012}: \textquotedblleft Schr\"{o}dinger's first thought was to attempt
to find a wave equation, moving on from de Broglie's work, that would
describe the behavior of an electron in the simplest atom, hydrogen. He
naturally included in his calculations allowance for the effects described
by the special theory of relativity, deriving, probably early in December
1925, what became known as the relativistic Schr\"{o}dinger (not
hydrogen---our correction!) equation. Unfortunately, it didn't work. The
predictions of the relativistic equation did not match up with observations
of real atoms. We now know that this is because Schr\"{o}dinger did not allow
for the quantum spin of the electron, which is hardly surprising, since the
idea of spin had not been introduced into quantum mechanics at that time.
But it is particularly worth taking note of this false start, since it
highlights the deep and muddy waters in which quantum physicists go
swimming---for you need to take account of spin, a property usually
associated with particles, in order to derive a wave equation for the
electron!%
\footnote{The concept of electron spin was introduced by G.~E.~Uhlenbeck and S.~Goudsmit in a
letter published in {\it{Die Naturwissenschaften\/}}; the issue of 20 November 1925; see \cite{Mehra1969} for more details\/.}

With the Christmas break coming up, he had an opportunity to get away from Z%
\"{u}rich and think things over in the clean air of Arosa %
...   \cite{Gribbin2012} \/.\textquotedblright (See also \cite{Mehra1987} and \cite{Moore1989} for more details%
{\footnote{Arosa, an Alpine \emph{Kurort} at about 1800~m altitude, not far from ski-resort Davos, and overlooked by the great peak of the Weisshorn.
For a related video, see: https://www.news.uzh.ch/en/articles/2017/Schroedinger.html \/. Here, among other things, a female physicist meets
the grandson of Dr.~Herwig and he shows the
entry of the payment in a guest book, done by Schr\"{o}dinger\/. 
}}%
.)

Whereas the honorable historians of science are arguing about the sequence and details of some events  \cite{Joas2009}, \cite{Kragh1982}, \cite{Mehra1987},
\cite{Wessels1979},
our remarks are dedicated to a somewhat missing, modern mathematical, part of this story, namely,
the discovery of a relativistic (spin-zero) wave equation by Erwin Schr\"{o}%
dinger (and why he did report and publish only the centennial
nonrelativistic version for the hydrogen atom).
\begin{figure}[hbt!]
\centering
\includegraphics[width=0.79\textwidth]{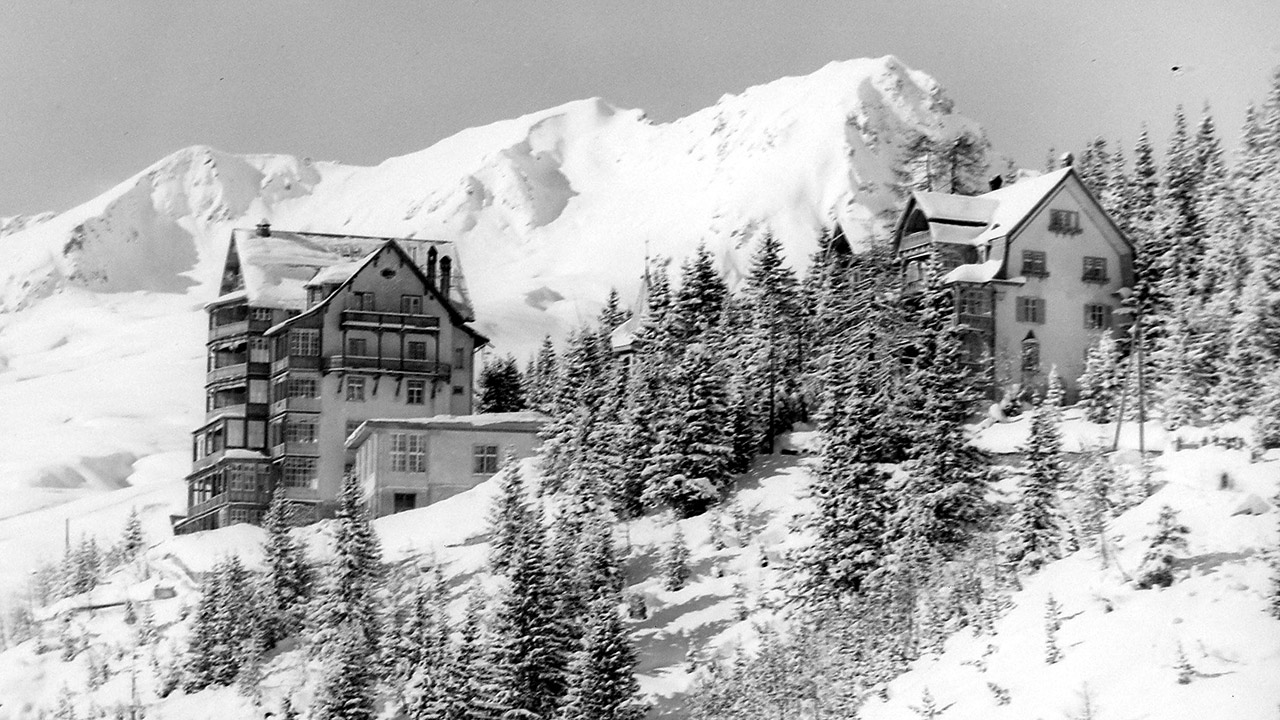}
\caption{The Villa Herwig--Frisia (right), Arosa, where it is believed wave mechanics was discovered during the Christmas holidays 1925--26\/.}
\label{Figure1}
\end{figure}
\begin{figure}[hbt!]
\centering
\includegraphics[width=0.67\textwidth]{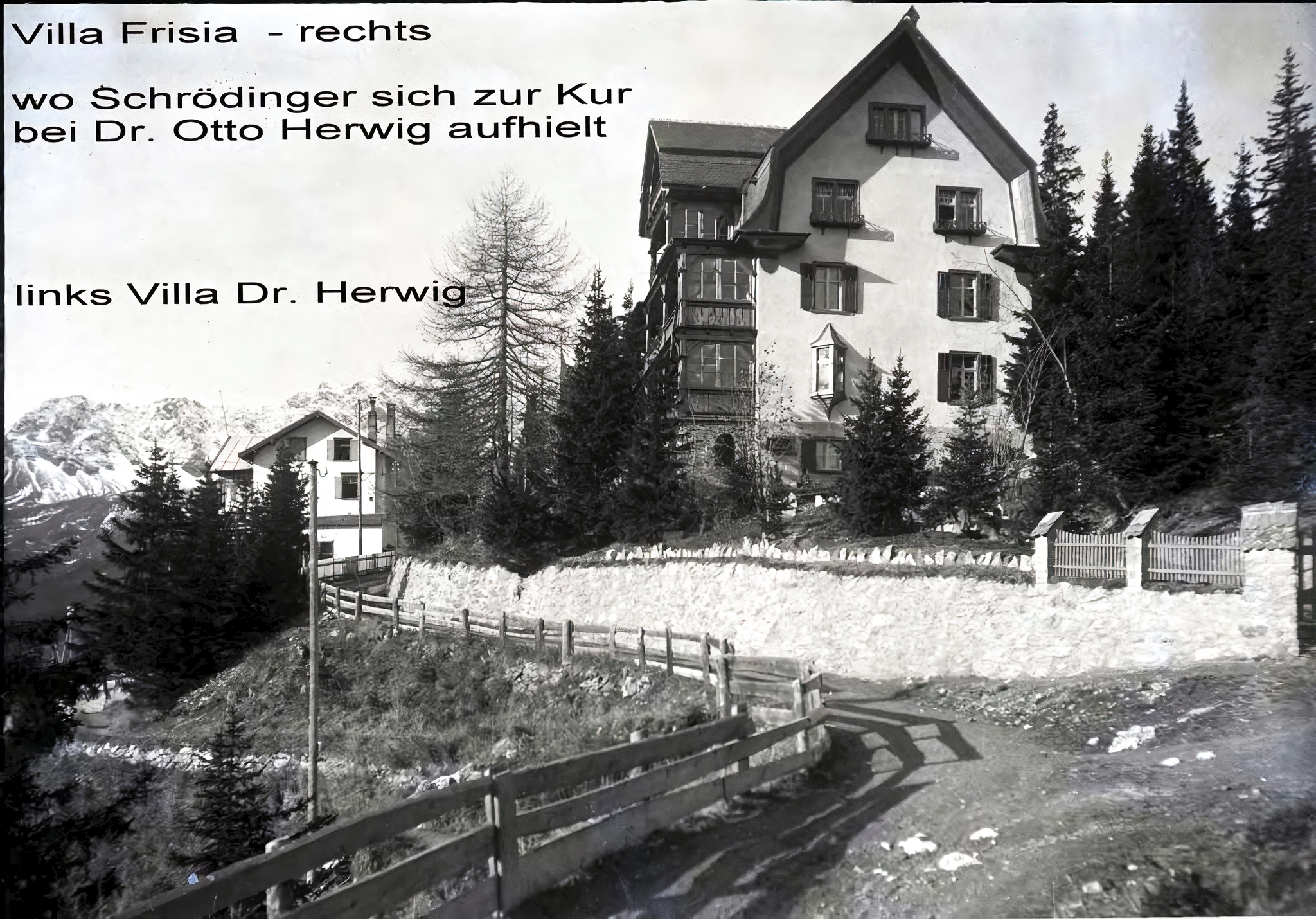}
\caption{The Villa Herwig--Frisia postcard.}
\label{Figure2}
\end{figure}

\section{Introducing the relativistic Schr\"{o}dinger equation\/}

How Schr\"{o}dinger derived his celebrated equation and subsequently applied it to the hydrogen atom?
According to his own testimony \cite{Schrodinger2010}, \cite{Schrodinger1926}, de Broglie's seminal work on a wave theory of matter of 1923--24
and Einstein's work on ideal Bose gases of 1924--25 laid the ground for the discovery of wave mechanics.
The crucial phase is dated approximately from late November 1925 to late January 1926 (see, for instance, \cite{Kragh1982} and \cite{Mehra1987}).

Let us start our discussion by following the original Schr\"{o}dinger's idea --- ``rumor has it''! For a classical relativistic particle
of charge $e$ in the electromagnetic field, one has to replace%
\begin{equation}
E^{2}=c^{2}\mathbf{p}^{2}+m^{2}c^{4}  \label{rel0}
\end{equation}%
by%
\begin{equation}
\left( E-e\varphi \right) ^{2}=\left( c\mathbf{p}-e\mathbf{A}\right)
^{2}+m^{2}c^{4} . \label{rel1}
\end{equation}%
In relativistic quantum mechanics, we replace the four-vector of energy $E$
and linear momentum $\mathbf{p}$ with the following operators%
\begin{equation}
E\rightarrow \widehat{E}=i\hbar \frac{\partial }{\partial t},\qquad \mathbf{p%
}\rightarrow \widehat{\mathbf{p}}=-i \hbar \nabla .  \label{rel3}
\end{equation}%
As a result, we arrive at the relativistic Schr\"{o}dinger equation \cite{Dirac1928}, \cite{Schiff1968}\/:%
\begin{align}
&\left( -\hbar ^{2}\frac{\partial ^{2}}{\partial t^{2}}-2ie\hbar \varphi
\frac{\partial }{\partial t} -ie\hbar \frac{\partial \varphi}{\partial t}+e^{2}\varphi ^{2}\right) \psi \\  \label{rel1a}
&\qquad=\left( -\hbar%
^{2}c^{2}\Delta +2ic\hbar \mathbf{A}\cdot \nabla + 2iec\hbar \, \nabla \cdot \mathbf{A} + e^2 {\mathbf{A}}^2  +m^{2}c^{4}\right) \psi \notag \/  .
\end{align}%
In the special case of Coulomb field\/,%
\begin{equation}
e\varphi =-\frac{Ze^{2}}{r},\qquad \mathbf{A}=\mathbf{0},  \label{rel4}
\end{equation}%
one can separate the time and spatial variables%
\begin{equation}
\psi \left( \mathbf{r},t\right) =\chi (\mathbf{r})e^{-i(E\;t)/\hbar }.
\label{rel5}
\end{equation}%
Finally, the stationary relativistic Schr\"{o}dinger equation takes the form
\begin{equation}
\left( E+\frac{Ze^{2}}{r}\right) ^{2}\chi =\left( -\hbar ^{2}c^{2}\Delta
+m^{2}c^{4}\right) \chi .  \label{rel6}
\end{equation}%
(As one can see, it is NOT a traditional eigenvalue problem!) The standard rumor says
that Schr\"{o}dinger invented this equation sometime in late 1925, found
the corresponding relativistic energy levels, and derived the fine structure
formula, different from Sommerfeld's one \cite{Dirac1963}. This is why he had
to abandon this equation and to consider a nonrelativistic approximation. Moreover,
they say that he had to withdraw the `relativistically framed' article, whose draft
never been found, from a journal and start all over with his centennial article on the nonrelativistic Schr%
\"{o}dinger equation \cite{Schrodinger2010}. Our goal here is to follow the
original `relativistic idea'. Our analysis, nonetheless, will suggest that it is hardly possible to believe
that a complete relativistic solution (for a charged spin-zero particle) had been found at that time
(see, nonetheless, \cite{Mehra1987} for an opposite opinion and a remarkable article of Vladimir~A.~Fock \cite{Fock1926}
for the first published solution of this problem among other things; see also \cite{Fock1926a}).

\section{Solving the relativistic Schr\"{o}dinger equation\/}

We follow \cite{Nikiforov1988} with somewhat different details (see also \cite%
{Davydov1965}, \cite{Fock1926}, \cite{Schiff1968} for a different approach) and, first of all,
separate the variables in spherical coordinates, $\chi \left( r,\theta
,\varphi \right) =R(r)Y_{lm}(\theta ,\varphi ),$ where $Y_{lm}(\theta
,\varphi )$ are the spherical harmonics with familiar properties \cite%
{Varshalovich1988}.\ As a result,%
\begin{equation}
\frac{1}{r^{2}}\frac{d}{dr}\left( r^{2}\frac{dR}{dr}\right) +\left[ \frac{%
\left( E+Ze^{2}/r\right) ^{2}-m^{2}c^{4}}{\hbar ^{2}c^{2}}-\frac{l\left(
l+1\right) }{r^{2}}\right] R=0\qquad (l=0,1,2,\;...)  \label{sol0}
\end{equation}%
\begin{figure}[hbt!]
\centering
\includegraphics[width=0.77\textwidth]{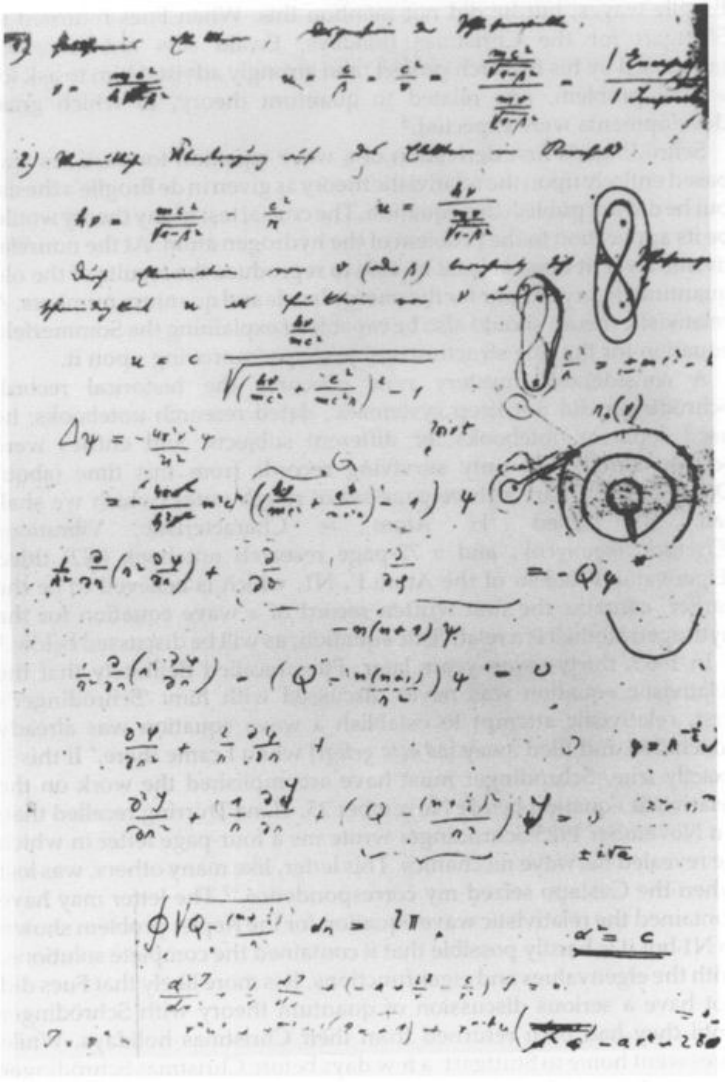}
\caption{A page from Notebook $N1\/$, with the first record of the wave equation \cite{Kragh1982}, \cite{Moore1989} (probably written about Christmas 1925\/).}
\label{Figure3}
\end{figure}
(see Figure~3 for the original version%
\footnote{Schr\"{o}dinger's notebooks are reproduced in the Archive for the History of Quantum Physics (AHQP);
for more details, see  \cite{Joas2009}, \cite{Mehra1987}\/.}
). In the dimensionless quantities,%
\begin{equation}
\varepsilon =\frac{E}{mc^{2}},\qquad x=\beta r=\frac{mc}{\hbar }r,\qquad \mu
=\frac{Ze^{2}}{\hbar c},  \label{sol1}
\end{equation}%
for the new radial function,%
\begin{equation}
R(r)=F(x)=\frac{u(x)}{x},  \label{sol2}
\end{equation}%
one gets%
\begin{equation}
\frac{1}{x^{2}}\frac{d}{dx}\left( x^{2}\frac{dF}{dx}\right) +\left[ \left(
\varepsilon +\frac{\mu }{x}\right) ^{2}-1-\frac{l\left( l+1\right) }{x^{2}}%
\right] F=0.  \label{sol3}
\end{equation}%
Given the identity,%
\begin{equation}
\frac{1}{x^{2}}\frac{d}{dx}\left( x^{2}\frac{dF}{dx}\right) =\frac{1}{x}%
\frac{d^{2}}{dx^{2}}\left( xF\right) ,  \label{sol4}
\end{equation}%
we finally obtain%
\begin{equation}
u^{\prime \prime }+\left[ \left( \varepsilon +\frac{\mu }{x}\right) ^{2}-1-%
\frac{l\left( l+1\right) }{x^{2}}\right] u=0.  \label{sol5}
\end{equation}%
This is the generalized equation of hypergeometric form (\ref{A1}) below,
when%
\begin{equation}
\sigma \left( x\right) =x,\qquad \widetilde{\tau }(x)\equiv 0,\qquad
\widetilde{\sigma }(x)=\left( \varepsilon ^{2}-1\right) x^{2}+2\mu
\varepsilon x+\mu ^{2}-l(l+1)  \label{sol6}
\end{equation}%
(see also Table~2 provided in Appendix~A below).

In view of the normalization condition,%
\begin{equation}
\int_{0}^{\infty }R^{2}(r)r^{2}\ dr=1,\qquad \text{or\quad }\int_{0}^{\infty
}u^{2}(x)\;dx=\beta ^{3},\quad \beta =\frac{mc}{\hbar ^{2}},  \label{sol7}
\end{equation}%
it's useful to start from asymptotic as $r\rightarrow 0$ $($or $x\rightarrow
0).$ In this limit, from (\ref{sol5}), one gets Euler's equation:%
\begin{equation}
x^{2}u^{\prime \prime }+\left[ \mu ^{2}-l\left( l+1\right) \right] u=0,
\label{sol8}
\end{equation}%
which can be solved by the standard substitution $u=x^{\alpha },$ resulting
in the following quadratic equation:%
\begin{equation}
\alpha ^{2}-\alpha +\left[ \mu ^{2}-l\left( l+1\right) \right] =0.
\label{sol9}
\end{equation}%
Thus the general solution of Euler's equation has the form%
\begin{equation}
u=C_{1}x^{\nu +1}+C_{2}x^{-\nu };\qquad \nu =-\frac{1}{2}+\sqrt{\left( l+%
\frac{1}{2}\right) ^{2}-\mu ^{2}},\quad \mu =\frac{Ze^{2}}{\hbar c};
\label{sol10}
\end{equation}%
which is bounded as $x\rightarrow 0$ with $C_{2}=0:$ $\ u\simeq C_{1}x^{\nu
+1}\/.$%
\footnote{The substitution $u=x^{\nu+1}U$ in (\ref{sol5}) results in Laplace's equation of the form:%
$$
xU^{\prime \prime}+2(\nu+1)U^{\prime}+\left(2\varepsilon\mu-a^2 x\right)U=0\/ \quad \left(a=\sqrt{1-\varepsilon^2}\right)\/,
$$
which can be solved by a complex integration \cite{Landau1998}, \cite{Schlesinger1900}; see also Appendix~C for some details. According to \cite{Mehra1987},
Schr\"{o}dinger did analyze these integral solutions in his three page Notebook $N1$! (As the first step towards solution of the nonrelativistic problem in Arosa?)
It was not realized at that time that a simple substitution $U=e^{-ax}y$ results in the Sturm--Liouville eigenvalue problem for the Laguerre polynomials!
(Thus the quantization rule is related to the number of zeros of the radial wave functions.)}

We continue our analysis of Schr\"{o}dinger's equation (\ref{sol5}), applying Nikiforov--Uvarov's technique, and transform (\ref{A1}) into the simpler form (\ref{A3}) by using $u\left(
x\right) =\varphi \left( x\right) y\left( x\right) ,$ where%
\begin{equation}
\frac{\varphi ^{\prime }}{\varphi }=\frac{\pi \left( x\right) }{\sigma
\left( x\right) },\qquad \pi (x)=\frac{\sigma ^{\prime }-\widetilde{\tau }}{2%
}\pm \sqrt{\left( \frac{\sigma ^{\prime }-\widetilde{\tau }}{2}\right) ^{2}-%
\widetilde{\sigma }+k\sigma },  \label{sol11}
\end{equation}%
or, in this particular case,%
\begin{eqnarray}
\pi (x) &=&\frac{1}{2}\pm \sqrt{\left( l+\frac{1}{2}\right) ^{2}-\mu
^{2}-2\mu \varepsilon x+\left( 1-\varepsilon ^{2}\right) x^{2}+kx}
\label{sol12} \\
&=&\frac{1}{2}\pm \left\{
\begin{array}{cc}
\sqrt{1-\varepsilon ^{2}}\;x+\nu +\frac{1}{2}, & k=2\mu \varepsilon +(2\nu
+1)\sqrt{1-\varepsilon ^{2}}; \\
\sqrt{1-\varepsilon ^{2}}\;x-\nu -\frac{1}{2}, & k=2\mu \varepsilon -(2\nu
+1)\sqrt{1-\varepsilon ^{2}}.%
\end{array}%
\right.  \notag
\end{eqnarray}%
Here, one has to choose the second possibility, when $\tau ^{\prime }=2\pi
^{\prime }<0,$ namely$:$%
\begin{equation}
\tau \left( x\right) =2\left( \nu +1-ax\right) ,\qquad a=\sqrt{1-\varepsilon
^{2}}.  \label{sol13}
\end{equation}%
Thus%
\begin{eqnarray}
\pi \left( x\right) &=&\nu +1-ax,\qquad \qquad \quad \varphi \left( x\right)
=x^{\nu +1}e^{-ax};  \label{sol14} \\
\lambda &=&2\left( \mu \varepsilon -(\nu +1)a\right) ,\qquad \rho \left(
x\right) =x^{2\nu +1}e^{-2ax}.  \notag
\end{eqnarray}%
For example,%
\begin{eqnarray*}
\left( \ln \varphi \right) ^{\prime } &=&\frac{\nu +1}{x}-a,\qquad \ln
\varphi =\left( \nu +1\right) \ln x-ax,\qquad \text{etc.} \\
\lambda &=&k+\pi ^{\prime }=2\left( \mu \varepsilon -(\nu +1)a\right) .
\end{eqnarray*}%
(All these data are collected in Table~2, Appendix~A, for the reader's convenience.)

In Nikiforov--Uvarov's method, the energy levels can be obtained from the
quantization rule \cite{Nikiforov1988}:%
%
\begin{equation}
\lambda +n\tau ^{\prime }+\frac{1}{2}n\left( n-1\right) \sigma ^{\prime
\prime }=0\qquad (n=0,1,2,\ ...),  \label{sol15}
\end{equation}%
which gives Schr\"{o}dinger's fine structure formula (for a charged
spin-zero particle in the Coulomb field) \cite{Fock1926}:%
\begin{equation}
E=E_{n_{r}}=\frac{mc^{2}}{\sqrt{1+\left( \dfrac{\mu }{n_{r}+\nu +1}\right)
^{2}}}\qquad \left( n=n_{r}=0,1,2,\ ...\right) .  \label{sol16}
\end{equation}%
Here,%
\begin{equation}
\mu =\frac{Ze^{2}}{\hbar c},\qquad \nu =-\frac{1}{2}+\sqrt{\left( l+\frac{1}{%
2}\right) ^{2}-\mu ^{2}}.  \label{sol17}
\end{equation}

The corresponding eigenfunctions are given by the Rodrigues-type formula%
\begin{equation}
y_{n}\left( x\right) =\frac{B_{n}}{x^{2\nu +1}e^{-2ax}}\frac{d^{n}}{dx^{n}}%
\left( x^{n+2\nu +1}e^{-2ax}\right) =C_{n}L_{n}^{2\nu +1}\left( 2ax\right) .
\label{sol18}
\end{equation}%
Up to a constant, they are Laguerre polynomials \cite{Nikiforov1988}. By (\ref%
{sol7}),%
\begin{eqnarray}
\beta ^{3} &=&\int_{0}^{\infty }u^{2}(x)\;dx=C_{n}^{2}\int_{0}^{\infty }
\left[ \varphi \left( x\right) L_{n}^{2\nu +1}\left( 2ax\right) \right]
^{2}\ dx  \notag \\
&=&\frac{C_{n}^{2}}{\left( 2a\right) ^{2\nu +3}}\int_{0}^{\infty }\xi ^{2\nu
+2}\left( L_{n}^{2\nu +1}\left( \xi \right) \right) ^{2}\ d\xi ,\quad \xi
=2ax.  \label{sol19}
\end{eqnarray}%
The corresponding integral is given by \cite{Suslov2020}:%
\begin{equation}
I_{1}=\int_{0}^{\infty }e^{-x}x^{\alpha +1}\left( L_{n}^{\alpha }\left(
x\right) \right) ^{2}\ dx=\left( \alpha +2n+1\right) \frac{\Gamma \left(
\alpha +n+1\right) }{n!}.  \label{sol20}
\end{equation}%
As a result,%
\begin{equation}
C_{n}=2\left( a\beta \right) ^{3/2}\left( 2a\right) ^{\nu }\sqrt{\frac{n!}{%
\Gamma \left( 2\nu +n+2\right) }}.  \label{sol21}
\end{equation}%
The normalized radial eigenfunctions, corresponding to the relativistic energy levels (%
\ref{sol16}), are explicitly given by%
\begin{equation}
R\left( r\right) =R_{n_{r}}\left( r\right) =2\left( a\beta \right) ^{3/2}%
\sqrt{\frac{n_{r}!}{\Gamma \left( 2\nu +n_{r}+2\right) }}\;\xi ^{\nu
}e^{-\xi /2}\;L_{n}^{2\nu +1}\left( \xi \right) ,  \label{sol22}
\end{equation}%
where%
\begin{equation}
\xi =2ax=2a\beta r=\sqrt{1-\varepsilon ^{2}}\frac{mc}{\hbar }r.
\label{sol23}
\end{equation}%
Our consideration here, or somewhat similar to that in \cite{Davydov1965}, \cite{Fock1926}, \cite%
{Schiff1968}, outlines the way which Schr\"{o}dinger had to complete (during his
Christmas vacation in 1925--26?) to discover the above solution of the
relativistic spin-zero charged particle in the central Coulomb field. One
can see that it's a bit more involved than the corresponding nonrelativistic case
originally published in Ref.~\cite{Schrodinger2010} (see also \cite%
{Bethe1957}, \cite{Landau1998}, \cite{Nikiforov1988}, \cite{Suslov2020} for modern treatments).

In his letter to Wilhelm Wien from Arosa (\cite{Meyenn2011}, pp.~162--165), Schr\"{o}dinger admits that working on ``a new atomic theory" he has ``... to learn mathematics
to handle the vibration problem ..." (see also \cite{Mehra1987a}, p.~775). In the first `quantization' article \cite{Schrodinger2010}, the Laplace method based on complex integration was utilized to solve the radial equation (for a nonrelativistic hydrogen atom). At that time, Schr\"{o}dinger refers to book \cite{Schlesinger1900}, well-known to him from his student days, and acknowledges some guidance from Hermann Weyl.
Nowadays, we may say that the corresponding quantization rule for those complex integrals is not that straightforward and a bit complicated --- it has hardly ever been utilized in the aforementioned form after that and now is completely forgotten.

Later on, Schr\"{o}dinger always uses (and develops) methods from the `magnificent classical mathematics' book \cite{Courant1924} (in Erwin's own words \cite{Mehra1987}, pp.~582--583) first published in 1924.
It's almost certain that this book was not available to him in Arosa.
Only in the second article on wave mechanics, Schr\"{o}dinger thanks his assistant E.~Fues for pointing out a connection with the Hermite polynomials for the harmonic oscillator problem and acknowledges the relation of his wave function in the `Kepler problem' with the `polynomial of Laguerre' \cite{Schrodinger2010ab}.

\section{Further treatment: nonrelativistic approximation\/}

According to Dirac \cite{Dirac1963}, \cite{Dirac1979} (see also \cite{Bloch1976}, \cite{Gribbin2012}, and \cite{Mehra1987}),
Schr\"{o}dinger was able to solve his relativistic wave equation (and submitted an
article to a journal, which original manuscript had never been found?). But later he had to
abandon the whole idea after discovering an inconsistency of his fine
structure formula with Sommerfeld's one, which was in agreement with the
available experimental data \cite{Sommerfeld1951}. During his Christmas 1925-vacation at Arosa, Schr%
\"{o}dinger introduced, eventually, his famous nonrelativistic stationary wave
equation and solved it for the hydrogen atom \cite{Schrodinger2010} (thus ignoring the fine structure).
The
relativistic Schr\"{o}dinger equation for a free spin-zero particle was
later rediscovered by Klein, Fock, and Gordon.%
\footnote{According to \cite{Kragh1982}, Louis de Broglie discovered the relativistic wave equation for a spin-zero free particle on February 1925 \cite{Broglie1925}, which did not attract any interest at that time\/.}
The relativistic wave
equation for a free electron, spin-$1/2$ particle, was discovered by Dirac only in 1928 \cite{Dirac1928}.%
\footnote{For a definition of the spin, in terms of representations of the Poincar\'{e} group, see \cite{Kryuchkov2016}\/.}
Its
solution in the central Coulomb field, when Sommerfeld--Dirac's fine structure
formula had been finally obtained from relativistic quantum mechanics, was later found
by Darwin \cite{Darwin1928} and Gordon \cite{Gordon1928}. As a result, the relativistic energy levels of an
electron in the central Coulomb field are given by%
\begin{equation}
E=E_{n_{r},\, j}=\frac{mc^{2}}{\sqrt{1+\mu ^{2}/\left( n_{r}+\nu \right) ^{2}}}%
,\quad \mu =\frac{Ze^{2}}{\hbar c}\quad (n_{r}=0,1,2,\ ...) . \label{lim0}
\end{equation}%
In Dirac's theory,%
\begin{equation}
\nu =\nu_{\text{Dirac}} =\sqrt{\left( j+1/2\right) ^{2}-\mu ^{2}},  \label{lim1}
\end{equation}%
where $j=1/2,3/2,5/2,\ ...\ $ is the total angular momentum including the
spin of the relativistic electron. A detailed solution of this problem,
including the nonrelativistic limit, can be found in \cite{Nikiforov1988}, \cite%
{Suslov2020} (following Nikiforov--Uvarov's paradigm), or in classical sources
\cite{Bethe1957}, \cite{Davydov1965}, \cite{Fock1978}, \cite{Schiff1968}.

Let us analyze a nonrelativistic limit of Schr\"{o}dinger's fine structure
formula (\ref{sol16})--(\ref{sol17}) and compare with the corresponding
results for Sommerfeld--Dirac's one (\ref{lim0})--(\ref{lim1}). For the
relativistic Schr\"{o}dinger equation, in the limit $c\rightarrow \infty $
(or $\mu \rightarrow 0),$ one gets \cite{Davydov1965}, \cite{Schiff1968}:%
\begin{eqnarray}
\frac{E_{n_{r},\, l}}{mc^{2}} &=&\frac{1}{\sqrt{1+\dfrac{\mu ^{2}}{\left(
n_{r}+\frac{1}{2}+\sqrt{\left( l+\frac{1}{2}\right) ^{2}-\mu ^{2}}\right) }}}
\notag \\
&=&1-\frac{\mu ^{2}}{2n^{2}}-\frac{\mu ^{4}}{2n^{4}}\left( \frac{n}{l+1/2}-%
\frac{3}{4}\right) +\text{O}\left( \mu ^{6}\right) ,\quad \mu \rightarrow 0,
\label{lim2}
\end{eqnarray}%
which can be derived by a direct Taylor's expansion and/or verified by a
computer algebra system. Here, $n=n_{r}+l+1$ is the corresponding
nonrelativistic principal quantum number. The first term in this expansion
is simply the rest mass energy $E_{0}=mc^{2}$ of the charged spin-zero
particle, the second term coincides with the energy eigenvalue in the
nonrelativistic Schr\"{o}dinger theory and the third term gives the
so-called fine structure of the energy levels, which removes the degeneracy between
states of the same $n$ and different $l\/.$
Unfortunately,  Schr\"{o}dinger's
relativistic approach didn't account correctly for the fine structure of hydrogenlike
atoms, such as hydrogen, ionized helium, doubly-ionized lithium, etc.
For example, the total splitting of the fine-structure levels for the principal quantum number $n=2$ became $8/3$ times as large as the one
obtained from Sommerfeld's theory, which was in perfect agreement with experiments, notably for ionized helium
\cite{Davydov1965}, \cite{Mehra1987}, \cite{Schiff1968}, \cite{Sommerfeld1951}.%
\footnote{The maximum spreads of the fine-structure levels occur when $l=0\/,$ $l=n-1$ and $j=1/2\/,$ $j=n-1/2$ for (\ref{lim2}) and (\ref{lim3}), respectively
\cite{Davydov1965}, \cite{Schiff1968}. Therefore, for the quotient, one gets
$$
\frac{\Delta E_{\text{Schr\"{o}dinger}}}{\Delta E_{\text{Sommerfeld}}}=\frac{4n}{2n-1} \qquad (n=2,3,\, ...)\/.
$$}
This is why, it is almost certain that Schr\"{o}dinger, who always insisted on a very good fit of experimental data by the theoretical description,
would never consider publishing a `wrong theory' of the relativistic hydrogen atom!

In Dirac's theory of the relativistic electron, the corresponding limit has the
form \cite{Bethe1957}, \cite{Davydov1965}, \cite{Schiff1968}, \cite{Suslov2020}:%
\begin{equation}
\frac{E_{n_{r},\, j}}{mc^{2}}= 1-\frac{\mu ^{2}}{2n^{2}}-\frac{\mu ^{4}}{%
2n^{4}}\left( \frac{n}{j+1/2}-\frac{3}{4}\right) +\text{O}\left( \mu
^{6}\right) ,\quad \mu \rightarrow 0,  \label{lim3}
\end{equation}%
where $n=n_{r}+j+1/2$ is the principal quantum number of the nonrelativistic
hydrogenlike atom. Once again, the first term in this expansion is the rest
mass energy of the relativistic electron, the second term coincides with the
energy eigenvalue in the nonrelativistic Schr\"{o}dinger theory and the
third term gives the so-called fine structure of the energy levels --- the
correction obtained for the energy in the Pauli approximation which includes
the interaction of the spin of the electron with its orbital angular momentum.
The total spread in the energy of the fine-structure levels is in agreement with experiments
(see \cite{Bethe1957}, \cite{Davydov1965}, \cite{Fock1978}, \cite{Schiff1968},
\cite{Sommerfeld1951} for further discussion
of the hydrogenlike energy levels including comparison with the experimental
data). For an alternative discussion of the discovery of Sommerfeld's formula,
see \cite{Granovskii2004}, \cite{Petrov2020}.

In a handwritten supplement to the February 20th, 1926 letter to Sommerfeld (\cite{Meyenn2011}, p.~181--182), Schr\"{o}dinger had analyzed introduction of magnetic field into his relativistic wave equation (see also \cite{Davydov1965} for the nonrelatiistic limit).

Let us also consider a nonrelativistic limit of the radial wave functions (%
\ref{sol22})--(\ref{sol23}) of the relativistic Schr\"{o}dinger equation.
From (\ref{sol17}), $\nu \rightarrow l$ as $c\rightarrow \infty \/.$ Moreover,%
\begin{equation}
1-\varepsilon ^{2}=\left( 1-\varepsilon \right) (1+\varepsilon )\rightarrow
\frac{\mu ^{2}}{2n^{2}}\cdot 2=\frac{\mu ^{2}}{n^{2}},\quad \mu \rightarrow
0,  \label{lim4}
\end{equation}%
and, therefore,%
\begin{equation}
\xi =\sqrt{1-\varepsilon ^{2}}\ \frac{mc}{\hbar }r\rightarrow 2\frac{\mu }{n}%
\cdot \frac{mc}{\hbar }r=\frac{2Z}{n}\left( \frac{r}{a_{0}}\right) =\eta
,\quad \mu \rightarrow 0,  \label{lim4a}
\end{equation}%
where $a_{0}=\hbar ^{2}/me^{2}$ is Bohr's radius. As a result, from (\ref%
{sol22})--(\ref{sol23}) one gets the nonrelativistic radial wavefunctions of
the hydrogenlike atom \cite{Bethe1957}, \cite{Landau1998}, \cite%
{Schrodinger2010}, \cite{Suslov2020}:%
\begin{equation}
R\left( r\right) \rightarrow R_{nl}\left( r\right) =\frac{2}{n^{2}}\left(
\frac{Z}{a_{0}}\right) ^{3/2}\sqrt{\frac{\left( n-l-1\right) !}{\left(
n+l\right) !}}\ \eta ^{l}\ e^{-\eta /2}\ L_{n-l-1}^{2l+1}\left( \eta \right)
\label{lim5}
\end{equation}%
with%
\begin{equation}
\eta =\frac{2Z}{n}\left( \frac{r}{a_{0}}\right) ,\qquad a_{0}=\frac{\hbar
^{2}}{me^{2}}\quad \text{(Bohr's radius)}  \label{lim5a}
\end{equation}%
and%
\begin{equation*}
\int_{0}^{\infty }R_{nl}^{2}\left( r\right) \ dr=1.
\end{equation*}%
In Table~1, provided in Appendix~A, we present the corresponding data for the nonrelativistic
hydrogenlike atom for the reader's convenience. (More details can be found
in \cite{Nikiforov1988}, \cite{Suslov2020}.)

Most likely that Schr\"{o}dinger was unaware of de~Broglie's relativistic wave equation of February 1925, which according to \cite{Kragh1982} was never mentioned by any physicist in 1925--27. Moreover, it's almost certain that this equation doesn't seem to have had any impact on his route towards the nonrelativistic eigenvalue problem (stationary Schr\"{o}dinger equation; see also \cite{Darrigol2013}). In a letter to Hermann Weyl written on April 1st, 1931 (see \cite{Meyenn2011}, pp.~483--485), regarding the second German edition of his book on group theory and quantum mechanics \cite{Weyl1931}, Schr\"{o}dinger describes his own contribution versus de Broglie's as follows: ``Now one more thing, please do not resent me. You have titled the \S\S 2 and 3 in Chapter II with the names of de Broglie and Schr\"{o}dinger. It looks as if the first of them knew about the assignment of the operators and the wave equation in field-free space, but not about the curvature of the rays in the potential field; and as if only the latter had taken this influence of the field into account by introducing the potentials into the wave equation. I don't think that's consistent.
If you are of the opinion that de Broglie's considerations about beam path and wavelength are actually a description of the wave equation (even if he does not use them in algebraic symbols written down), then you should put the wave equation including potentials under `de Broglie'.
Then he was not one iota less aware of the behavior in the field than of the behavior in the field-free space. Otherwise, however, it is really a bit annoying to read, under the de Broglie name, for example, the operator mapping  ($p_1 \to {\partial}/{\partial q_1}$), which as far as I know is actually announced for the first time in my paper from March 18th, 1926 (see \cite{Schrodinger2010ac}, p.~47), and also the scalar relativistic equation, which I described on the first page of my original paper \cite{Schrodinger2010} (of course only in words) and the solutions of which I published in \S 3, section 2 of the same paper.
I never protested the fact that this equation is now commonly known under the name of Gordon, because this is convenient for distinguishing it [from non-relativistic one].
But for obvious reasons, I don't like to see it under the name de Broglie (for example, on p.~211 of your book)
[page 188 in the second German edition]."

Our translation of the entire letter from German to English, that seems is missing in the available literature, can be found in Appendix~D.
Here, Schr\"{o}dinger's somewhat emotional response may be related to a certain Nobel Prize controversy.
Louis de~Broglie was awarded the prize in 1929 after discovery of electron-diffraction picture as an experimental confirmation of his idea of the wave nature of electrons.
At that time, the Nobel committee was very reluctant to give the prize for a pure theoretical work because, due to the will of Alfred Nobel, the prize should be given to the person whose `discovery or invention' shall have `conferred the greatest benefit on mankind'. In the next two years the prizes in physics were not awarded for the discovery of quantum mechanics although both Heisenberg and Schr\"{o}dinger were nominated multiple times. (The 1930 prize went to Raman `for his work on the scattering of light and for the discovery of the effect named after him' and no prize was awarded in 1931 after vigorous debates in the committee and the international physics community.) Only after experimental discovery of the positron in cosmic rays, The Nobel Prize in Physics 1932 was given to Heisenberg `for the creation of quantum mechanics, the application of which has, inter alia, led to the discovery of the allotropic forms of hydrogen' and The Nobel Prize for 1933 was divided between Schr\"{o}dinger and Dirac `for the discovery of new productive forms of atomic theory' \cite{Moore1989}.

\section{Semiclassical approximation: WKB-method for Coulomb fields\/}

Phenomenological quantization rules of `old' quantum mechanics \cite{Sommerfeld1951} are derived in modern physics
from the corresponding wave equations in the so-called semiclassical approximation \cite{Blokhintsev1964}, \cite{Davydov1965}, \cite{Landau1998} (WKB-method).
In this approximation, for a particle in the central field, one can use a generic radial equation of the form \cite{Langer1937}, \cite{Nikiforov1988}:
\begin{equation}
u''(x)+ q(x)\, u=0\/, \label{wkb5.1}
\end{equation}%
where $x^2\, q(x)$ is continuous together with its first and second derivatives for $0\leq x\leq b <\infty\/.$
As is well-known, the traditional semiclassical approximation cannot be used in a neighborhood of $x=0\/.$
However, the change of variables $x=e^z\/$, $u=e^{z/2}\, v(z)$ transforms the equation into a new form,
\begin{equation}
v''(z)+ q_1(z)\, v=0\/, \label{wkb5.2}
\end{equation}%
where
\begin{equation}
q_1(z)=-\dfrac{1}{4} + \left( x^2 \, q(x)\right)_{x=e^z} \/ \label{wkb5.3}
\end{equation}%
(Langer's transformation \cite{Langer1937}). As $z\to -\infty$ (or $x\to 0$), the new function $q_1(z)$ is changing slowly near the constant $-1/4 + \lim_{x\to 0}x^2q(x)\/$ and
$\lim_{z\to -\infty}q_1^{(k)}(z)=0\/$ ($k=1,2$). Hence the function $q_1(z)$ and its derivatives are changing slowly for large negative $z\/.$
The WKB-method can be applied to (\ref{wkb5.2}) and, as a result, in the original equation (\ref{wkb5.1}) one should replace $q(x)$ with $q(x)-1/(4x^2)\/$
(see \cite{Langer1937}, \cite{Nikiforov1988} for more details).

For example, in the well-known case of nonrelativistic Coulomb's problem, one gets
\begin{equation}
u^{\prime \prime }+\left[ 2 \left( \varepsilon_0 +\frac{Z }{x}\right)-%
\frac{\left( l+1/2\right)^2 }{x^{2}}\right] u=0 \quad \left(\varepsilon_0 = \dfrac{E}{E_0}, \, \, \, E_0 = \dfrac{e^2}{a_0}, \, \, \, a_{0}=\frac{\hbar
^{2}}{me^{2}}\right) \/  \label{wkb5.4}
\end{equation}%
(in dimensional units, see Table~1 or \cite{Suslov2020}).
The Bohr--Sommerfeld quantization rule takes the form
\begin{equation}
\int_{r_1}^{r_2} p(r) \, dr = \pi \left( n_r + \dfrac{1}{2} \right) \qquad \qquad (n_r=0\/, 1\/, 2\/, \, \dots)  \label{wkb5.5}
\end{equation}%
with
\begin{equation}
p(r)= \left[ 2 \left( \varepsilon_0 +\frac{Z }{r}\right)-%
\frac{\left( l+1/2\right)^2 }{r^{2}}\right]^{1/2} , \qquad p(r_1)=p(r_2)=0\/  \label{wkb5.6}
\end{equation}%
(see, for instance, \cite{Nikiforov1988} for more details).

For all Coulomb problems under consideration, we utilize the following generic integral, originally evaluated by Sommerfeld \cite{Sommerfeld1951}
performing complex integration: If
\begin{equation}
p(r)= \sqrt{-A +\dfrac{B}{r}-\dfrac{C}{r^2}}\;\qquad A\/,C>0 \/ . \label{wkb5.6}
\end{equation}%
Then
\begin{equation}
\int_{r_1}^{r_2} p(r) \, dr = \pi \left( \dfrac{B}{2\sqrt{A}} - \sqrt{C} \right)  \label{wkb5.7}
\end{equation}%
provided $p(r_1)=p(r_2)=0\/.$ (See Appendix~B for an elementary evaluation of this integral.)

In (\ref{wkb5.4}), $A=-2\varepsilon_0\/,$ $B=2Z\/,$ and $C=(l+1/2)^2.\/$  In view of the quantization rule (\ref{wkb5.5}) ,
\begin{equation}
\dfrac{Z}{\sqrt{-2\varepsilon_0}} - l - \dfrac{1}{2} = n_r  +  \dfrac{1}{2} \/ . \label{wkb5.8}
\end{equation}%
As a result, we obtain exact energy levels for the nonrelativistic hydrogenlike problem:
\begin{equation}
\varepsilon_0 = \dfrac{E}{E_0} = - \dfrac{Z^2}{2(n_r +l+1)^2}\/ . \label{wkb5.9}
\end{equation}
Here, $n=n_r +l +1$ is known as the principal quantum number.

Our main goal is to analyze the corresponding relativistic problems. In the case of the relativistic Schr\"{o}dinger equation (\ref{sol5})--(\ref{sol6}),
one should write
\begin{equation}
u^{\prime \prime }+\left[ \left( \varepsilon +\frac{\mu }{x}\right) ^{2}-1-%
\frac{\left( l+1/2 \right)^2 }{x^{2}}\right] u=0  \label{wkb5.10}
\end{equation}%
and
\begin{equation}
p(x)= \left[ \left( \varepsilon +\frac{\mu }{x}\right) ^{2}-1-%
\frac{\left( l+1/2 \right)^2 }{x^{2}}\right]^{1/2}\/ . \label{wkb5.11}
\end{equation}
Here, $A=1-{\varepsilon}^2\/,$ $B= 2\mu \varepsilon\/,$ and $C=(l+1/2)^2-{\mu}^2\/.$ The quantization rule (\ref{wkb5.5}) implies
\begin{equation}
\dfrac{\mu \varepsilon}{\sqrt{1- {\varepsilon}^2}} = n_r +\nu +1 \/ \label{wkb5.12}
\end{equation}
and the exact formula for the relativistic energy levels (\ref{sol16})--(\ref{sol17}) follows once again.%
\footnote{This is, almost certain, how Schr\"{o}dinger was able to derive his fine structure formula so quickly, from `old' quantum mechanics,
and verify the result with the help of his new `complex' quantization rule. But
a true reason why this algebraic equation, obtained here in semiclassical approximation, coincides with the exact one
in relativistic quantum mechanics still remains a mystery!\/}

Sommerfeld's fine structure formula (\ref{lim0})--(\ref{lim1}), for the relativistic Coulomb problem, can be thought of as the main achievement of the `old' quantum mechanics \cite{Granovskii2004}, \cite{Milantev2004}, \cite{Sommerfeld1951}. Here,
we will derive this result in a semiclassical approximation for the radial Dirac equations (separation of variables in spherical coordinates is discussed in detail in Refs.~\cite{Nikiforov1988} and \cite{Suslov2020}).
In the dimensionless units, one of these second-order differential equations has the form
\begin{equation}
v''_1 + \dfrac{({\varepsilon}^2-1)x^2 +2\varepsilon \mu x - \nu(\nu+1)}{x^2} \, v_1 = 0 \/ \label{wkb5.13}
\end{equation}
and the second equation can be obtained from the first one by replacing $\nu \to -\nu\/$ (see Eqs.~(3.81)--(3.82) in Ref.~\cite{Suslov2020}).
By Langer's transformation, we obtain
\begin{equation}
p(x)= \left[ \left( \varepsilon +\frac{\mu }{x}\right) ^{2}-1-%
\frac{\left( \nu+1/2 \right)^2 + \mu^2}{x^{2}}\right]^{1/2}\/ . \label{wkb5.14}
\end{equation}
Hence, for the Dirac equation,  $A=1-{\varepsilon}^2\/,$ $B= 2\mu \varepsilon\/,$ and $C=(\nu+1/2)^2.\/$
The quantization rule (\ref{wkb5.5}) implies (\ref{wkb5.12}). (One has to replace $n_r \to n_r - 1\/$, see \cite{Suslov2020} for more details.)
Thus the fine structure formula for the energy levels (\ref{lim0})--(\ref{lim1}) has been derived here in semiclassical approximation
for the corresponding radial Dirac equations. This explains the quantization rules of the `old' quantum mechanics
of Bohr and Sommerfeld \cite{Sommerfeld1951} (see also \cite{Granovskii2004}). The integral (\ref{wkb5.6})--(\ref{wkb5.7}) was well-known to
Schr\"{o}dinger --- it is mentioned in his late January 1926 letter to Sommerfeld when he
reported on the first success of the wave mechanics (see \cite{Mehra1987}, p.~462 or \cite{Meyenn2011}, p.~172).
%

%

\section{Erwin Schr\"{o}dinger --- 60 years after\/}

In saved correspondence and numerous working notes, Schr\"{o}dinger never gave a chronological pathway to the birth of wave mechanics
that made him one of the classics of science.
Schr\"{o}dinger's personal diaries for the years 1925--26 are not available.
This is why we may never know for sure all the details of his revolutionary discovery.
Nonetheless, our analysis of the relativistic Schr\"{o}dinger equation allows suggesting that he, probably, didn't
have a chance to solve this equation for Coulomb potential, say from the viewpoint of the later standards, when eigenvalues and
the normalized eigenfunctions should be explicitly presented?  (The first solution was published by V.~A.~Fock just a few months later \cite{Fock1926}).
Instead, he seems found the corresponding fine structure
formula for a charged spin-zero particle by using the Bohr--Sommerfeld quantization rule and/or by the Laplace method (correcting a mistake in the original
work \cite{Granovskii2004}, \cite{Joas2009}?) and observed discrepancies with the experimental data \cite{Mehra1987}.
Analyzing this `disaster', he switched to the nonrelativistic case and introduced the stationary Schr\"{o}dinger
equation, which he was able to solve around December-January of 1925--26 (by the Laplace method, consulting with Hermann Weyl?).
His main legacy, namely, the time-dependent Schr\"{o}dinger equation (Figure~4),
{\it{de facto}} required in the article dedicated to the coherent states \cite{Schrodinger2010a},%
\footnote{The coherent states were introduced as a result of correspondence with Hendrik Lorentz \cite{Kox2013}, \cite{Moore1989};
see also \cite{Kryuchkov2013} for an extension to the minimum-uncertainty squeezed states.\/}
was published only about six months later \cite{Schrodinger2010b}. His review \cite{Schrodinger1926} gives an account of the 
new form of quantum theory. That article was submitted on September~3, 1926. At the end, addressing the controversy with the relativistic hydrogen atom, Schr\"{o}dinger concludes: ``The deficiency must be intimately connected with Uhlenbeck--Goudsmit's
theory of the spinning electron. But in what way the electron
spin has to be taken into account in the present theory is yet unknown".
Less than two years later, Paul Dirac would derive a first-order wave equation for a four-component spinor field that describes relativistic spin-1/2 particles
like electrons \cite{Dirac1928}.

The year 1926 revolutionary changed the world of physics forever --- the Schr\"{o}dinger equation turned out to have an enormously wide range of applications \cite{Miller2002}, from the quantum theory of atoms and molecules to solid state physics, quantum crystals, superfluidity, and superconductivity.
In a historical perspective, the `golden years' following the foundation of quantum mechanics have paved the way to quantum field theory and the physics of elementary particles.
Just in a few months, Erwin Schr\"{o}dinger turned the matter wave hypothesis of Louis de Broglie into a detailed, perfectly operating technique, which eventually eliminated many of `old' quantum physics postulates, that had been employed in the previous decades, as unnecessary.
However, at that time, a `true nature' of quantum motion, including the probabilistic interpretation of the wave function, the uncertainty principle, and the wave-particle duality, was not yet well-understood --- it came later as a result of vigorous debates in the newly (re)born `quantum community' \cite{Mehra1987}. Some of these  debates are not completed even 60 years after Schr\"{o}dinger!
In mathematics, the time-dependent Schr\"{o}dinger equation gave rise to a new class of partial differential equations \cite{Dyson2009}.

In November 1926, Schr\"{o}dinger wrote in the preface to a collected edition of his first papers on wave mechanics \cite{Schrodinger2010c}:
``Referring to these six papers (the present reprint of which is solely due to the great demand
for separate copies), a young lady friend%
\footnote{The young friend was fourteen-year old Itha Junger, whom Schr\"{o}dinger was tutoring in mathematics \cite{Gribbin2012}, \cite{Moore1989}.}
recently remarked to the author: `When you began this work you had no idea that anything so clever would come out of it, had you?'
This remark, with which I wholeheartedly agreed
(with due qualification of the flattering adjective), may serve to call attention to the fact that the papers now combined
in one volume were originally written {\it{one by one\/}} at different times. The results of the later sections were largely
unknown to the writer of the earlier ones." No alterations of the originals were allowed. In the end, Schr\"{o}dinger concludes:
``On the contrary, this comparatively cheap method of issue seemed advisable on account of the impossibility at the present stage of giving a fresh exposition which would be really satisfactory or conclusive." Since then his classical works combined in one volume and translated to English remain a monumental `snapshot' in the history of science.
After Schr\"{o}dinger's review~\cite{Schrodinger1926} (immediately translated into Russian by Soviet Physics Uspekhi, 1927) and the subsequent blast of research, the first `really satisfactory and conclusive' expositions of the newly emerged quantum theory were given by Werner Heisenberg \cite{HeisenbergQM} (in German, 1930), Paul Dirac \cite{DiracPrinciples} (in English, 1930), and Vladimir~A.~Fock \cite{Fock1978} (in Russian, 1931). A detailed list of early sources can be found in \cite{Elyashevich1977} and \cite{Weyl1931}.

In a way, we have had somewhat similar `feelings' (with due qualification, of course!), originally writing these notes mostly for our colleagues, friends, and students and thoroughly checking the resulting interpretations with the available bibliography (letters, notebooks, and articles). We hope that this consideration, despite potential imperfections, will encourage the readers to study quantum physics at one of the crucial moments of its creation and draw their own conclusions.

\begin{figure}[hbt!]
\centering
\includegraphics[width=0.825\textwidth]{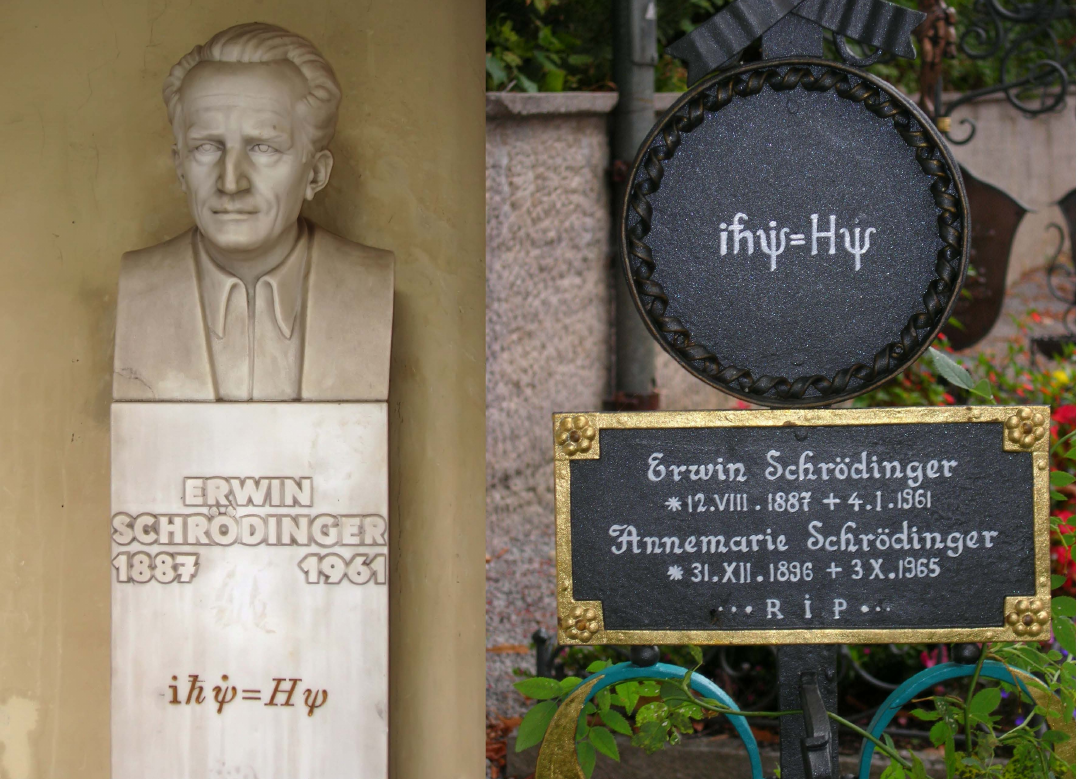}
\caption{The Schr\"{o}dinger equation --- University of Vienna (left) and Alpbach (right), respectively. }
\label{Figure4}
\end{figure}
\begin{appendix}
\appendix

\section{Summary of the Nikiforov--Uvarov method\/}

Generalized equation of the hypergeometric type%
\begin{equation}
u^{\prime \prime }+\frac{\widetilde{\tau }(x)}{\sigma (x)}u^{\prime }+\frac{%
\widetilde{\sigma }(x)}{\sigma ^{2}(x)}u=0  \label{A1}
\end{equation}%
($\sigma ,$ $\widetilde{\sigma }$ are polynomials of degrees at most $2, %
\widetilde{\tau }$ is a polynomial at most first degree) by the substitution%
\begin{equation}
u=\varphi (x)y(x)  \label{A2}
\end{equation}%
can be reduced to the form%
\begin{equation}
\sigma (x)y^{\prime \prime }+\tau (x)y^{\prime }+\lambda y=0  \label{A3}
\end{equation}%
if:%
\begin{equation}
\frac{\varphi ^{\prime }}{\varphi }=\frac{\pi (x)}{\sigma (x)},\qquad \pi
(x)=\frac{1}{2}\left( \tau (x)-\widetilde{\tau }(x)\right)  \label{A4}
\end{equation}%
(or, $\tau (x)=\widetilde{\tau }+2\pi ,$ for later),%
\begin{equation}
k=\lambda -\pi ^{\prime }(x)\qquad (\text{or, }\lambda =k+\pi ^{\prime }),
\label{A5}
\end{equation}%
and%
\begin{equation}
\pi (x)=\frac{\sigma ^{\prime }-\widetilde{\tau }}{2}\pm \sqrt{\left( \frac{%
\sigma ^{\prime }-\widetilde{\tau }}{2}\right) ^{2}-\widetilde{\sigma }%
+k\sigma }  \label{A6}
\end{equation}%
is a linear function. (Use the choice of constant $k$ to complete the square
under the radical sign; see \cite{Nikiforov1988} and our note below for more details.)

In Nikiforov--Uvarov's method, the energy levels can be obtained from the
quantization rule:%
\begin{equation}
\lambda +n\tau ^{\prime }+\frac{1}{2}n\left( n-1\right) \sigma ^{\prime
\prime }=0\qquad (n=0,1,2,\ ...)  \label{A7}
\end{equation}%
and the corresponding square-integrable solutions are classical orthogonal polynomials, up to a factor.
They can be found by the Rodrigues-type formula \cite{Nikiforov1988}:
\begin{equation}
y_n(x) = \dfrac{B_n}{\rho(x)} \left(\sigma^n(x) \rho(x)\right)^{(n)}, \qquad (\sigma \rho)'=\tau \rho , \label{A8}
\end{equation}%
where $B_n$ is a constant (see also \cite{Suslov2020}).

The corresponding data for the nonrelativistic and relativistic hydrogenlike problems  are presented in Tables~1 and 2, respectively.


%
\begin{table}[h!]
\centering
\begin{tabular}{|l|l|}
\hline
$\sigma (x)$ & $x$ \\ \hline
$\widetilde{\sigma }(x)$ & $2\varepsilon_0 \;x^{2}+2Z\;x-l\left( l+1\right)\/, \quad {\varepsilon_0 =E/E_0 } $
\\ \hline
$\widetilde{\tau }(x)$ & $0$ \\ \hline
$k$ & $2Z-\left( 2l+1\right) \sqrt{-2\varepsilon_0\/}$ \\ \hline
$\pi \left( x\right) $ & $l+1-\sqrt{-2\varepsilon_0 }\;x$ \\ \hline
$\tau \left( x\right) =\widetilde{\tau }+2\pi \qquad $ & $2\left( l+1-\sqrt{%
-2\varepsilon_0 }\;x\right) $ \\ \hline
$\lambda =k+\pi ^{\prime }$ & $2\left( Z-(l+1)\sqrt{-2\varepsilon_0 }\right) $
\\ \hline
$\varphi \left( x\right) $ & $x^{l+1}e^{-x\sqrt{-2\varepsilon_0 }}$ \\ \hline
$\rho (x)$ & $x^{2l+1}e^{-(2Z\ x)/n},\quad x=r/a_{0}$ \\ \hline
$y_{n_{r}}(x)$ & $C_{n_{r}}L_{n_{r}}^{2l+1}\left( \eta \right) ,\quad \eta =%
\dfrac{2Z}{n}x=\dfrac{2Z}{n}\left( \dfrac{r}{a_{0}}\right) $ \\ \hline
$C_{n_{r}}^{2}$ & $\dfrac{Z}{a_{0}n^{2}}\dfrac{\left( n-l-1\right) !}{\left(
n+l\right) !},\quad n_{r}=n-l-1$ \\ \hline
\end{tabular}
\caption[]{Schr\"{o}dinger equation for the Coulomb potential $V(r)=-\dfrac{%
Ze^{2}}{r}.$ \\
 Dimensionless quantities:\\
  \begin{minipage}{\linewidth}
\begin{equation*}
r=a_{0}x,\quad a_{0}=\frac{\hbar ^{2}}{m_{e}e^{2}}\simeq 0.5\cdot 10^{-8}\;%
\text{cm},\quad E_{0}=\frac{e^{2}}{a_{0}};\quad R\left( r\right) =F\left(
x\right) =\frac{u\left( x\right) }{x}.
\end{equation*}
  \end{minipage}}
\end{table}
%
%


%
%
\begin{table}[h!]
\centering
\begin{tabular}{|l|l|}
\hline
$\sigma (x)$ & $x$ \\ \hline
$\widetilde{\sigma }(x)$ & $\left( \varepsilon ^{2}-1\right) \;x^{2}+2\mu
\varepsilon \;x+\mu ^{2}-l\left( l+1\right) $ \\ \hline
$\widetilde{\tau }(x)$ & $0$ \\ \hline
$k$ & $2\mu \varepsilon -\left( 2\nu +1\right) \sqrt{1-\varepsilon ^{2}}%
,\quad \nu =-\dfrac{1}{2}+\sqrt{\left( l+\dfrac{1}{2}\right) ^{2}-\mu ^{2}}$
\\ \hline
$\pi \left( x\right) $ & $\nu +1-a\;x,\qquad a=\sqrt{1-\varepsilon ^{2}}$ \\
\hline
$\tau \left( x\right) =\widetilde{\tau }+2\pi $ & $2\left( \nu
+1-a\;x\right) $ \\ \hline
$\lambda =k+\pi ^{\prime }$ & $2\left( \mu \varepsilon -(\nu +1)a\right) $
\\ \hline
$\varphi \left( x\right) $ & $x^{\nu +1}e^{-a\ x}$ \\ \hline
$\rho (x)$ & $x^{2\nu +1}e^{-2a\ x}$ \\ \hline
$y_{n_{r}}(x)$ & $C_{n_{r}}L_{n_{r}}^{2\nu +1}\left( \xi \right) ,\quad \xi
=2ax=2a\beta r=2\sqrt{1-\varepsilon ^{2}}\dfrac{mc}{\hbar }r$ \\ \hline
$C_{n_{r}}$ & $2\left( a\beta \right) ^{3/2}\left( 2a\right) ^{\nu }\sqrt{%
\dfrac{n_{r}!}{(\nu +n_{r}+1)\Gamma \left( 2\nu +n_{r}+2\right) }}$ \\ \hline
\end{tabular}
\caption[]{Relativistic Schr\"{o}dinger equation for the Coulomb potential $%
V(r)=-\dfrac{Ze^{2}}{r}.$  \\
Dimensionless quantities:\\
  \begin{minipage}{\linewidth}
\begin{equation*}
\varepsilon =\frac{E}{mc^{2}},\quad x=\beta r=\frac{mc}{\hbar }r,\quad \mu =%
\frac{Ze^{2}}{\hbar c};\qquad R\left( r\right) =F\left( x\right) =\frac{%
u\left( x\right) }{x}.
\end{equation*}
  \end{minipage}}
\end{table}
%
%


\noindent \textbf{Note.\/} A closed form for the constant $k$ can be obtained as follows.
Let
\begin{equation}
p(x)={\left( \frac{%
\sigma ^{\prime }-\widetilde{\tau }}{2}\right) ^{2}-\widetilde{\sigma }%
+k\sigma } =q(x) + k\sigma(x)\/,  \label{A9}
\end{equation}
where
\begin{equation}
q(x)=\left(\frac{\sigma ^{\prime }-\widetilde{\tau }}{2}\right)^2 -\widetilde{\sigma } \/. \label{A10}
\end{equation}
Completing the square, one gets
\begin{equation}
p(x)=\dfrac{p^{\prime \prime }}{2}\left(x+\frac{p^{\prime}(0)}{p^{\prime \prime }}\right)^2 -\dfrac{\left(p^{\prime}(0)\right)^2-2p^{\prime \prime }p(0)}{2p^{\prime \prime }} \/, \label{A11}
\end{equation}
where the last term must be eliminated:%
\begin{equation}
{\left(p^{\prime}(0)\right)^2-2p^{\prime \prime }p(0)} = 0 \/. \label{A12}
\end{equation}
Therefore,
\begin{equation}
\left(q^{\prime}(0)+k\sigma^{\prime} (0)\right)^2-2\left(q^{\prime \prime } + k \sigma ^{\prime \prime } \right) \left(q(0) + k\sigma(0) \right) = 0 \/, \label{A13}
\end{equation}
which results in the following quadratic equation:
\begin{equation}
ak^2+2bk+c=0. \label{A14}
\end{equation}
Here,
\begin{align}
&a = \left(\sigma^{\prime}(0) \right)^2  - 2\sigma^{\prime \prime} \sigma(0) \/, \\ 
&b = q^{\prime}(0) \sigma^{\prime}(0) - \sigma^{\prime \prime} q(0) -\sigma(0) q^{\prime \prime} \/,   \\ 
&c =   \left(q^{\prime}(0) \right)^2  - 2q^{\prime \prime} q(0) \/. 
\end{align}
Solutions are
\begin{equation}
k_0 = -\dfrac{c}{2b}, \qquad \text{if} \quad a=0 \label{A18}
\end{equation}
and
\begin{equation}
k_{1,2}=\dfrac{-b \pm \sqrt{d}}{a}, \qquad \text{if} \quad a\ne 0 \/. \label{A19}
\end{equation}
Here,
\begin{equation}
d=b^2-ac=\left(\sigma(0)q^{\prime \prime} + \sigma^{\prime \prime} q(0) \right)^2 - 2\left(\sigma^{\prime}(0)q^{\prime \prime} - \sigma^{\prime \prime} q^{\prime}(0) \right)
\left(\sigma(0)q^{\prime }(0) - \sigma^{\prime }(0)q(0) \right) \/. \label{A20}
\end{equation}
As a result,
\begin{equation}
p(x)=\left(\dfrac{p^{\prime \prime }}{2}\left(x+\frac{p^{\prime}(0)}{p^{\prime \prime }}\right)^2 \right)_{k=k_{0, 1, 2}} \/, \label{A21}
\end{equation}
which allows evaluating linear function $\pi(x)\/$ in the Nikiforov--Uvarov technique. (Examples are presented in Tables~1 and 2.)

\section{Evaluation of the integral\/}

Integrating by parts on the right-hand side of (\ref{wkb5.7}), one gets
\begin{align}
&{ \left. \int_{r_1}^{r_2} p(r) \, dr =   r p(r) \right\vert}_{r_1}^{r_2} - \int_{r_1}^{r_2} \dfrac{r\left[-(B/r^2) + 2(C/r^3) \right]}{2p(r)}\, dr \label{B1} \\
&\quad = \dfrac{B}{2} \int_{r_1}^{r_2} \dfrac{dr}{\sqrt{-A r^2 + B r - C}} - \int_{r_1}^{r_2} \dfrac{(C/r^2)\, dr }{\sqrt{-A + (B/r) - (C/r^2)}} \/. \notag
\end{align}%
For the last but one integral, we can write
\begin{equation}
\int_{r_1}^{r_2} \dfrac{dr}{\sqrt{\left(\frac{B^2}{4{A}}-C\right)-\left(r \sqrt{A} - \frac{B}{2\sqrt{A}} \right)^2}} \/
= {\left.
\dfrac{1}{\sqrt{A}}\arcsin  \dfrac{r \sqrt{A} - \frac{B}{2\sqrt{A}}}{{\sqrt{\frac{B^2}{4A}-C}}}
\right\vert}_{r_1}^{r_2} = \dfrac{\pi}{\sqrt{A}}\ .
\label{B2}
\end{equation}
The following substitution $r=1/x\/,$ in the last integral of (\ref{B1}), results in
\begin{equation}
-\int_{x_1=1/{r_1}}^{x_2=1/{r_2}} \dfrac{C \, dx}{\sqrt{-A+Bx-Cx^2}} = -\sqrt{C} \int_{x_1}^{x_2} \dfrac{\sqrt{C} \, dx}{\sqrt{\left(\frac{B^2}{4C}-A\right)-\left(x\sqrt{C}- \frac{B}{2\sqrt{C}}\right)^2}} = \pi \, \sqrt{C}\/ ,
\end{equation}
where we have completed the square and performed a familiar integral evaluation. (One may also interchange $A$ and $C$ in the previous integral evaluation.) Combining the last two integrals, one can complete the proof.

\section{The Laplace method\/}

The second order ordinary differential equation with linear coefficients,
\begin{equation}
L[u]:=x\dfrac{d^2 u}{dx^2}+\left(\delta_0 x + \delta_1\right) \dfrac{du}{dx} + \left(\varepsilon_0 x + \varepsilon_1\right) u=0 \/, \label{C1}
\end{equation}
the so-called Laplace equation, is easy to handle for the following reason: the Laplace transform gives an equation of the first order \cite{Schlesinger1900},  \cite{Schrodinger2010}.
Letting
\begin{equation}
u(x)=\int_C e^{xz} v(z)\, dz \label{C2}
\end{equation}
one gets
\begin{align}
&xu''+\delta_0 x u' + \varepsilon_0 x u=x \int_C e^{xz} \left[\left(z^2 + \delta_0 z + \varepsilon_0\right)v(z)\right]\, dz  \notag \\
&\quad = \int_C \dfrac{d}{dz} \left(e^{xz}\right) \left[\left(z^2 + \delta_0 z + \varepsilon_0\right)v(z)\right]\, dz  \notag \\
&\quad =\int_C \dfrac{d}{dz} \left[ e^{xz} \left(z^2 + \delta_0 z + \varepsilon_0\right)v(z)\right]\, dz  \notag \\
&\qquad - \int_C e^{xz}  \dfrac{d}{dz} \left[\left(z^2 + \delta_0 z + \varepsilon_0\right)v(z)\right]\, dz  \label{C3}
\end{align}
and
\begin{equation}
\delta_1 u' + \varepsilon_1 u = \int_C e^{xz} \left[\left(\delta_1 z + \varepsilon_1\right)v(z)\right] \, dz \/ .
\label{C4}
\end{equation}
Thus
\begin{align}%
&L[u](x) =\int_C \dfrac{d}{dz} \left[ e^{xz} \left(z^2 + \delta_0 z + \varepsilon_0\right)v(z)\right]\, dz  \notag \\
&\qquad\quad -\int_C e^{xz}  \dfrac{d}{dz} \left[\left(z^2 + \delta_0 z + \varepsilon_0\right)v(z)\right]\, dz \notag \\
&\qquad\qquad + \int_C e^{xz} \left[\left(\delta_1 z + \varepsilon_1\right)v(z)\right] \, dz  \equiv 0\/
\label{C5}
\end{align}
provided
\begin{equation}
\dfrac{d}{dz} \left[\left(z^2 + \delta_0 z + \varepsilon_0 \right) v(z) \right]= \left(\delta_1 z + \varepsilon_1\right)v(z)
\label{C6}
\end{equation}
and the corresponding boundary condition holds.

Let $c_1$ and $c_2$ be roots of the following quadratic equation:
\begin{equation}
z^2 + \delta_0 z + \varepsilon_0 = (z-c_1) (z-c_2) = 0\/ .
\label{C7}
\end{equation}
Denoting $v_1(z) = (z-c_1) (z-c_2) v(z)\/,$ we obtain
\begin{equation}
\dfrac{v_1^{'}(z)}{v_1(z)}= \dfrac{\delta_1 z + \varepsilon_1}{(z-c_1) (z-c_2)} = \dfrac{\alpha_1}{z-c_1} + \dfrac{\alpha_2}{z-c_2} \/,
\label{C8}
\end{equation}
where
\begin{equation}
\alpha_1 = \dfrac{c_1\delta_1 + \varepsilon_1}{c_1-c_2} \/, \qquad \alpha_2 = \dfrac{c_2\delta_1 + \varepsilon_1}{c_2-c_1}   \/.
\label{C9}
\end{equation}
As a result, one gets
\begin{equation}
v(z) =(z-c_1)^{\alpha_1 -1} (z-c_2)^{\alpha_2 -1} \/
\label{C10}
\end{equation}
and our solution take the form
\begin{equation}
u(x)=\int_C e^{xz} (z-c_1)^{\alpha_1 -1} (z-c_2)^{\alpha_2 -1} \, dz
\label{C11}
\end{equation}
provided that the following holds for the contour $C$:
\begin{equation}
\int_C \dfrac{d}{dz} \left[ e^{xz} (z-c_1)^{\alpha_1} (z-c_2)^{\alpha_2} \right] \, dz = 0 \/ .
\label{C12}
\end{equation}
For original applications of this method to the Coulomb problems see \cite{Schrodinger2010} and \cite{Fock1926}; more historical details are available in \cite{Mehra1987}, pp.~426--434.

According to the original Schr\"{o}dinger rudimentary quantization rule, the exponents ${\alpha_1}$ or ${\alpha_2}$ in the above complex integral solution were required to be negative integers. In his letter to Sommerfeld, Schr\"{o}dinger wrote with excitement that ``Finally I wish to add that the discovery of the whole connection [between the wave equation and the quantization of hydrogen atom], goes back to your beautiful quantization method for evaluating the radial quantum integral. ... which suddenly shone out from the exponents  ${\alpha_1}$ and ${\alpha_2}$ like a Holy Grail.'' (see \cite{Mehra1987}, p.~462 or \cite{Meyenn2011}, p.~172).

\section{A Letter from Schr\"{o}dinger to Weyl\/}

\noindent
Berlin-Grunewald, 1.~April~1931
\newline
\noindent
[Typescript]

\medskip
\noindent
Dear Mr.~Weyl,
\bigskip

\noindent
Once again, I have let pass a long period of time, in a completely unproportional way,
before hereby saying thanks for your prestigious, beautiful present -- the second
edition of your book. I have found most interesting the new presentation of Dirac's
theory since I am dealing with it currently.
As far as I can see, you have decided to use a new terminology since you consider the following as important:
to express that one can arrange that the Dirac matrices -- with the unique exception of the mass term
-- transform in such a way the first couple and the second couple of the components
of $\psi$ among themselves, and moreover, that this property also holds with respect to
an arbitrary Lorenz transform.
One can perceive (and you mention this somewhere in
the text) that your hope focusses on the fact that all the four components and
equations correspond to the electron and proton. I wish that this hope might become
fulfilled very soon. [The ``right" antiparticle -- positron will be discovered only next year!]

May I tell you about some vague thoughts which confirm my impressions about this?
If you consider the components of the ``current" as ``a probability for the electron
propagation process through the area element" then one can explicitly say that this is
-- similar to the gas kinetic diffusion theory -- a difference of the probability for passing
in one way or the other. Something similar then must hold for the density. Certain
points which can classically be considered as the trace of a world line, perforating the
volume element, must be considered as positive, others as negative.
This means the density must be in connection with the algebraic sum of charges. This fact must be
applicable to electrons as well as protons -- we should point out however that the
current density then has four perforation points -- what at first sight is somehow
perturbing -- or alternatively, one has to assume something like the theory of holes,
like it is done by Dirac.
Within such a theory, the world lines of protons are really in a
certain sense world lines which pass from the future back towards the past. Namely if
such a hole spatially propagates from $A$ to $B,$ then you will find the following situation:
before the process, there is nothing in $A$ but something in $B.$ Afterwards, there is
something in $A$ but nothing in $B.$ (Is one allowed to call this inverted time direction, or
is it simply inverted spatial direction?)

I have to say that I am quite astonished about the following: I have more difficulties to
read your written words than to follow your speech. I believe it is related to the fact
that mathematical objects and conclusions are much more apparent to you than to
many other people. Hence you eliminate steps in a chain of thoughts which would be
necessary for others to be included. Like in the prominent example: ``A man thinks that
men are dangerous." It would be disgusting to add here the following conclusion:
``Consequently, this man is dangerous". I think that your aesthetical perception -- which
would avoid and not allow such a conclusion -- is an obstacle for you, to summarize
with primitive words at the end again what has been derived and what has been
proven.

I am most thankful for the register of operations symbols and symbols with a fixed
meaning. This constitutes an enormous relaxation in contrast to the first edition.

Moreover I see that many, many things are new, being an enormous extension of the
first edition which remains in some aspects just not accessible to me, taking into
regards my abilities to understand it. In my brain, understanding quantum mechanics
takes place at such a slow speed -- and slower than in the case of any other man.
Ehrenfest's picture of a small asthmatic dachshund -- breathless following Electra --
can be applied to full extent to myself. Perhaps I will understand all the quantum facts
precisely at the moment when they will no more be valid.

One smaller aspect: on page 119, one finds the expression of intersection of two
subgroups. However this expression is not available in the index of the book. It is, as
far as I can see, for the first time present on page 105, but also there, it is not
explained explicitly.

A slight deviation, not following the main stream of the approach, I can moreover find
in comparing page~99, line~4 from above (``From now on ... only the one-to-one
maps.") and page~120, line~7 from above, where as far as I can see, a definition of
mapping is needed which is non unique.

I still have another point, and I hope you will not mind. As for the headline of
paragraphs~2 and 3 in chapter~2, you chose the names de Broglie and Schr\"{o}dinger. It
looks as if the first one knew about the assignment of operators and the wave
equation in a space without fields, but not about the curvature of rays in a potential
field. And moreover it looks as if it was the second one who has respected the
influence of the field by having introduced the potentials into the wave equation. I do
not think that this is a consequent way of argumentation. If you hold the view that the
approach by de Broglie with respect to ray propagation and wavelength constitute
already a description of the wave equation (even given the fact that he does not write
it down with algebraic symbols), then you will have to assign the wave equation,
including potentials, under the name ``de Broglie". The reason for this is the following:
the behavior of the objects in the field was known to him as well as the behavior of
the objects in a field-free space -- without a single jota difference in his perception.
On the other hand, it is indeed somewhat disturbing to find the operator assignment
$(p_1 \to \partial /\partial q_1)$ -- which to my knowledge has really been communicated for the first time in my manuscript
from 18th March 1926 -- being assigned to his name. The same perception applies to
the scalar relativistic equation which I have described on the first page of my first
treatment (although having used just written words).
About its solution, I have made some remarks in paragraph~3, and 2 of the same publication. I have never expressed
any protest against the fact that this equation now is correlated with the name
Gordon, since this helps determining the different scenarios. But I do not like to see
them being assigned to the name de Broglie -- for obvious reasons (compare for
instance also with page~188 of your book).

Please do not mind my objections, I hate deliberations of this type like the death, I
am not angry, nor offended or insulted. I know that something like this may happen if
one tries to order the thoughts behind given facts to one concluding, consequent,
compact stream of perception. While doing so, one has often to separate what is
united from the viewpoint of its historical origin. And on the other hand, hand one has
to unify what is historically separated. One should avoid telling: this one has made
this and that one has made that. In principle, all that is equivalent.
I have no doubt that you will understand my impression to a certain extent and you
will not decline its meaning to a certain extent.

And please, it would be horrible if you believe you would have to submit now
declarations and justifications. This is not my intention. I just wanted to talk about it. It
is not that important -- so let me end my deliberations there.

You are just returning from the USA, and hopefully you have had two pleasant and
not too stormy crossings overseas. I am little bit jealous about your travels overseas
(even in case they were stormy). But I am not too jealous about your stay in the US.
But just for 14~days, I would like to travel there if there is an opportunity, i.~e. if
someone else is going to pay for it. 14 days on a ship is the most beautiful and most
recovering Easter holiday -- well, it costs something.

\noindent
With the most cordial greetings and hand kisses from house to house \newline \noindent
Yours sincerely \qquad  \qquad Schr\"{o}dinger

\noindent \textbf{Acknowledgments.\/} We are grateful to Prof.~Dr.~Ruben Abagyan, Dr.~ Mark Faifman, Prof.~Dr.~ Ya\-kov~I.~Granovski\u{\i}, Prof.~Dr.~Georgy~T.~Guria,
Prof.~Dr.~Christian Krattenthaler, Dr.~Sergey~I.~Kryu\-chkov, Prof.~Dr.~Peter Paule, 
Dr.~Naya Smorodinskaya, Dr.~Eugene Stepanov,
and Prof.~Dr.~Alexei Zedanov for support, help, and valuable comments.
We are indebted to the referees, their comments helped us to improve the manuscript.
\end{appendix}

\providecommand{\bysame}{\leavevmode\hbox to3em{\hrulefill}\thinspace}
\providecommand{\noopsort}[1]{}
\providecommand{\mr}[1]{\href{http://www.ams.org/mathscinet-getitem?mr=#1}{MR~#1}}
\providecommand{\zbl}[1]{\href{http://www.zentralblatt-math.org/zmath/en/search/?q=an:#1}{Zbl~#1}}
\providecommand{\jfm}[1]{\href{http://www.emis.de/cgi-bin/JFM-item?#1}{JFM~#1}}
\providecommand{\arxiv}[1]{\href{http://www.arxiv.org/abs/#1}{arXiv~#1}}
\providecommand{\doi}[1]{DOI: \href{https://doi.org/#1}{#1}}
\providecommand{\MR}{\relax\ifhmode\unskip\space\fi MR }
\providecommand{\MRhref}[2]{%
  \href{http://www.ams.org/mathscinet-getitem?mr=#1}{#2}
}
\providecommand{\href}[2]{#2}

\end{document}